\documentclass[aps,prl,twocolumn,showpacs,preprintnumbers,amsmath,amssymb,groupedaddress]{revtex4-1}

\usepackage{soul}

\usepackage{graphicx}% Include figure files
\usepackage{dcolumn}% Align table columns on decimal point
\usepackage{bm}% bold math
\usepackage{color}
\definecolor{Red}  {rgb}{1,0,0}
\definecolor{Green}{rgb}{0,1,0}
\definecolor{Blue} {rgb}{0,0,1}
\newcommand {\bfv}[1] {{\boldsymbol {#1}}}

\newcommand\removed[1]{}

\newcommand\SEC[1]{\vskip 2mm {\it #1}.\rule[1mm]{5mm}{0.1mm}}
\usepackage{ulem}
\def\Skaku#1{{\color{Red}{#1}}}

\begin{document}

% Use the \preprint command to place your local institutional report
% number in the upper righthand corner of the title page in preprint mode.
% Multiple \preprint commands are allowed.
% Use the 'preprintnumbers' class option to override journal defaults
% to display numbers if necessary
%\preprint{}

%Title of paper
%\title{An Alternative Scinario of Laminar-Turbulent Transition in Planar Couette Flow $\sim$ Invisible Watershed on Transition}
%\title{The Role of the Hairpin Vortex Solution on Laminar-Turbulent Transition of Plane Couette Flow at Moderate Reynolds Number}
%\title{On the Laminar-Turbulent Transition of Plane Couette Flow at High Reynolds Number}
%\title{Structure of Turbulent Attraction of Plane Couette Flow at High Reynolds Number}
%\title{An Outline of the Basin of Turbulent Attraction of Plane Couette Flow at High Reynolds Number}
%\title{A Prediction on the Outline of the Basin of Turbulent Attraction of Plane Couette Flow at the infinite Reynolds Number}
%\title{\st{A} Transition \st{Scenario} of Plane Couette Flow at infinite Reynolds numbers - shall we remove 'scenario'?}
%\title{On the Outline of the Basin of Turbulent Attraction of Plane Couette Flow at the infinite Reynolds Number}
%\title{On an Ultimate structure of Plane Couette Flow at the infinite Reynolds Number}
\title{Transition of Planar Couette Flow at infinite Reynolds numbers}

% repeat the \author .. \affiliation  etc. as needed
% \email, \thanks, \homepage, \altaffiliation all apply to the current
% author. Explanatory text should go in the []'s, actual e-mail
% address or url should go in the {}'s for \email and \homepage.
% Please use the appropriate macro foreach each type of information
% \affiliation command applies to all authors since the last
% \affiliation command. The \affiliation command should follow the
% other information
% \affiliation can be followed by \email, \homepage, \thanks as well.
%\author{}
%\email[]{Your e-mail address}
%\homepage[]{Your web page}
%\thanks{}
%\altaffiliation{}
%\affiliation{}

\author{}
%\thanks{}
\affiliation{}

\author{Tomoaki Itano}
%\email{itano@ipcku.kansai-u.ac.jp}
\affiliation{
  Department of Pure and Applied Physics,
  Faculty of Engineering Science, Kansai University,  Osaka, 564-8680, Japan
}
\author{Takeshi Akinaga}
\thanks{Marie Curie Fellow}
\affiliation{
  School of Engineering and Applied Sciences, Division of Mathematics, Aston University, UK
}
\author{Sotos C. Generalis}
\affiliation{
  School of Engineering and Applied Sciences, Division of Mathematics, Aston University, UK
}
\author{Masako Sugihara-Seki}
\affiliation{
  Department of Pure and Applied Physics,
  Faculty of Engineering Science, Kansai University,  Osaka, 564-8680, Japan \ 
}

%Collaboration name if desired (requires use of superscriptaddress
%option in \documentclass). \noaffiliation is required (may also be
%used with the \author command).
%\collaboration can be followed by \email, \homepage, \thanks as well.
%\collaboration{}
%\noaffiliation

\date{\today}

\begin{abstract}
An outline of the state space of planar Couette flow at high Reynolds numbers ($Re < 10^5$) is investigated via a variety of efficient numerical techniques.
It is verified from nonlinear analysis that the lower branch of {\it Hairpin Vortex State} (HVS) \cite{itasot09,gib09} asymptotically approaches the primary (laminar) state with increasing $Re$.
%Moreover,  is emerged from a plenty of trajectories starting at initial conditions constituted by superposition of three distinct exact steady solutions of PCF, including the 
  It is also predicted that the lower branch of HVS at high $Re$ belongs to the stability boundary that initiates transition to turbulence,
  and that one of the unstable manifolds of the lower branch of HVS %, as well as HVS itself,  
  lies on the boundary.
These facts suggest HVS may provide a criterion to estimate a minimum perturbation arising transition to turbulent states at the infinite $Re$ limit. 
\end{abstract}

% insert suggested PACS numbers in braces on next line

% insert suggested keywords - APS authors don't need to do this
%\keywords{}

%\maketitle must follow title, authors, abstract, \pacs, and \keywords
\maketitle

%\section*{Introduction}
\SEC{Introduction}
%\paragraph{\bf Map of stars in space}
Astronomers have found a tremendous number of stars far away from the earth in attempting to map our universe, that is expanding due to the presence of the dark energy. Similarly, fluid physicists have recently elucidated that there exist a number of exact and invariant solutions in the state space of the turbulent shear flows governed by incompressible Navier-Stokes equations, that will map turbulence \cite{kaw12,len11}.
The first discovery of an invariant finite-amplitude solution distinct from the laminar state is the so-called Nagata's solution, referred to hereafter as ``NBW'' (formed from the names of the discoverers\cite{nag90,cle97,wal98}), was achieved in a simple shear flow between parallel plates moving in opposite directions, planar Couette flow (PCF).
NBW emerges at a rather lower $Re$ ($Re=127.7$) than the one that is required for sustaining the turbulent regime in PCF. %, where Reynolds number is defined as $Re=Uh/\nu$ in terms of the kinematic viscosity $\nu$, the channel half width $h$, and the moving-wall velocity $\pm U$.

%\paragraph{\bf NBW is {\color{Red} the} ultimate structure at $\bf Re\to \infty$   }
Wang {\it et al.} in Ref. \cite{wan07} recently calculated that the NBW, amongst a number of solutions, is not just an exact solution emerging at the relatively low $Re$, but also a significant \lq key\rq \ required to envisage the ideal turbulent regime at infinite $Re$ from the physical point of view, by pointing out the following facts.
Firstly, NBW remains on the {\it basin boundary} (BB) even at high $Re$.
BB is the hypersurface of  codimension-1 in the state space associated with PCF, which separates the whole space into the basins of the laminar and the turbulent attractions.
Secondly, NBW contains the only unstable manifold at high $Re$, which 
 %{\bf [(a) heads away from, (b) heads apart from, (c) heads  out of, (d) escapes from] } 
 escapes from the BB towards either the laminar or the turbulent attractions, so that NBW is an attractor of the dynamical system restricted on the BB.
These facts make us envisage that NBW could yield an ultimate structure prevalent in the turbulent regime at the infinite $Re$, even if it has an instability along the unstable manifold towards turbulent attraction.
It seems to be relevant to large-scale structures identified numerically in turbulent shear flows at high $Re$ \cite{del01,hwa11}.
%Based on these facts, they proposed a new possible approach to turbulence control method by utilizing the low-dimensionality of unstable manifolds of the lower branch of NBW.

%\paragraph{\bf Transition at high $\bf Re$}
By contrast, within a classical picture, hairpin-shaped vortices, with an $\Omega$-shaped head and a pair of counter-rotating streamwise legs, have been so far recognised as one of prevalent vortex structures in turbulent shear flows \cite{her88,rob91,adr07,wu09}. 
Here, we have to keep in mind that the NBW does not satisfy the reflection symmetry with respect to the spanwise direction, which is common in vortices observed in a transition triggered initially with a perturbation on the laminar flow.
The recent progress in numerical simulations is enabling us to educe vortex structures in turbulent boundary layers at high $Re$ over a huge domain \cite{sch10}, 
 % youtube upload / Marusic et al., Phys. Fluids, 2010 Klewicki, J. Fluids Eng., 2010 Smits et al., Annu. Rev. Fluid Mech., 2011, 
 in which neither NBW nor hairpin-shaped vortices are detected further downstream of the turbulent boundary layer.
However, in the turbulent layer, which develops with increasing $Re$ (based on the boundary thickness $Re_{\theta}$ \cite{sch10}), it is the hairpin-shaped vortices that emerge primarily at the beginning of the transition with a relatively high $Re$.

%\paragraph{\bf A contradiction}
Apparently, among these arguments there seems to be some fundamental contradiction on the envisaged ultimate structure in shear flow in the infinite $Re$; 
%An arising question is  ``what could we envisage as the ultimate coherent structure in the shear flows at the infinite Reynolds number limit ?''
 there arises a question ``why the envisaged ultimate structure, NBW, does not prevail at the transition with high $Re$ ?''
In the present study, we will give an answer to this question, with the aid of a transition scenario deduced from investigation of the asymptotic behaviour of another fundamental solution among a number of exact steady solutions of PCF, the {\it Hairpin Vortex State} (HVS) \cite{itasot09,gib09}.
%, with increase of Reynolds number.

%\section{Asymptotic behaviour of HVS}
\SEC{Asymptotic behaviour of HVS}
%By the way, Wang {\it et al.} in Ref. \cite{wan07} lately reported two key points on the lower branch of NBW, (1) that it exists on the {\it basin boundary} (BB), the hypersurface with codimension-1 in the phase space associated with PCF, which separates the whole phase space to the basins of the laminar and the turbulent attraction, and (2) that, moreover, it has the only unstable manifold,
 % which heads  \underline{[(1) away from, (2) apart from, (3) out of]}
% which escapes from the BB, so that it is an attractor of the dynamical system restricted on the BB at high Reynolds numbers.
%\paragraph{\bf HVS vs NBW}
%HVS was originally solved as a candidate of the exact solutions corresponding to hairpin-shaped vortex at relatively low Reynolds numbers $Re \sim 200$ \cite{itasot09,gib09}. %, where Reynolds number is defined as $Re=Uh/\nu$ in terms of the kinematic viscosity $\nu$, the channel half width $h$, and the moving-wall velocity $\pm U$.
%HVS was originally solved at relatively low Reynolds numbers $Re \sim 200$ \cite{itasot09,gib09}.
In the bifurcation diagram, HVS consists of the upper and lower branches arising from a saddle-node bifurcation at the turning point $Re=139.2$.
It should be again emphasised that the reflection symmetry to the spanwise direction is satisfied by HVS but not by NBW.
%By contrast, 
NBW emerges from a slightly lower $Re$ than HVS does, in spite of the fact that NBW bifurcates due to the breaking of the symmetry from HVS; in other words, NBW is a derivative (rather than the counterpart) of the HVS. 
%{\color{Red} Tomoaki-mou: This is just a repeat of the previous knowledge. I agree it is past, BUT IT NEEDS reminding here - so please leave it - do not take this paragraph away}.
%These solutions both exist futher lower than the lowest Reynolds number required for sustaining turbulent regime in PCF,  

%\paragraph{\bf Upper branch of HVS}
In particular, the upper branch of HVS yields hairpin-shaped vortex structures 
 %{\bf with the horizontally staggered formation (with the alignment of horiontally cross stripes ?) ( with the horizontal alignmnet of cross stripes ?)}
 aligned by stream- and span-wise staggered formations
 reminiscent to those seen in experiments, after which the solution was named. %, which are observed ubiquitously at transition stages in turbulent shear flows \cite{her88}.
%At a couple of legs of a vortex structure, the localised vorticity lift up the fluid near the boundary so as to form a couple of streaky regions near the boundary, and the head of the vortex structure induces coalescence of these streak structures, which is visualised as a bulge beneath the head of the vortex structure at $Re \le 300$.
%Though the upper branch of HVS is steady, nevertheless 
However, a radical change of the topology of its velocity field with increasing $Re$ ($Re>400$) gives rise to a complicated deformation of the vortex structure, which leads to enhancement of the fluid mixing between the moving plates, followed by steep rise of the wall shear rate. %requires high resolution in space to solve numerically.
While the upper branch of HVS has thus attracted much attention recently (\cite{sotita10}, \cite{deg10}) obviously owing to its characteristics such as spatial shape and staggered formation, the analysis of lower branch of HVS has been neglected in the recent focus of research.

%By contrast, 
%These solutions both exist futher lower than the lowest Reynolds number required for sustaining turbulent regime in PCF,  
%As a key in the transition at high Reynolds numbers, 
%another solution should be necessary to answer this fundamental contradiction, ``
%The vortex structures are sinusiodal in the 
%in turbulent shear flows
%which are observed ubiquitously 

%\paragraph{\bf Behaviour of NBW at the high Reynolds}
\begin{figure}
  \centerline{
    \includegraphics[angle=0,width=0.50\textwidth]{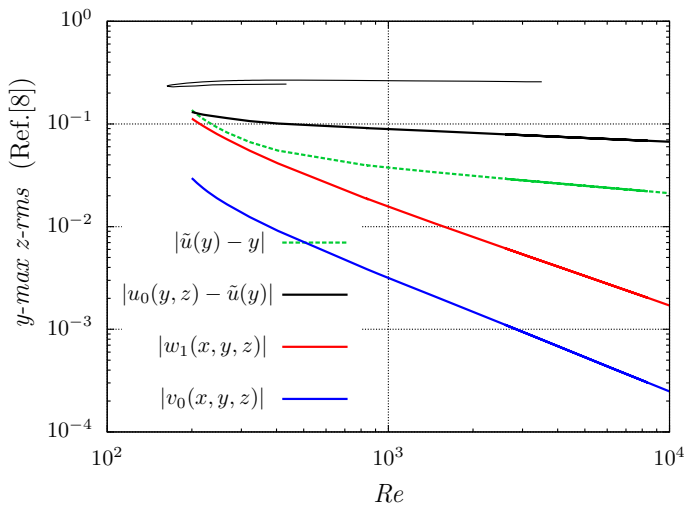}
  }
  \caption{
    Amplitudes of representative Fourier modes for HVS in PCF are plotted against $Re$ for stream- and span-wise extents of periodic box $(L_x,L_z)=(\pi,2\pi)$. 
    All the amplitudes of HVS algebraically decrease with increasing $Re$.
    In particular, gradual attenuation in streak component $|u_0(y,z)-\bar{u}(y)|$ of HVS (thick solid curve) contrasts well with that of NBW (thin solid curve), which is indicated to be kept at order 1  (as verified in Ref. \cite{wan07}).
    %This fact suggests the HVS, which could  bifurcate from infinity, may be a criterion of transition.
  }
  \label{figB}
\end{figure}
According to Wang {\it et al.}\cite{wan07}, the lower branch of NBW has an asymptotic behaviour at high $Re$ limit, which consists both of a non-vanishing streak component and of vanishing streamwise vortices.
Here, the ``streak component'' is referred to as the streamwise-independent perturbation $u_0(y,z)$ from the mean component $\bar{u}(y)$ of PCF, which is maintained by the streamwise vortices of NBW developing into a kind of critical layer at high $Re$. 
%This vortex sheet is a kind of critical layer getting to be thinner as increase of Reynolds number.
This fact means that the distance of NBW from the laminar state in the state space, $||\bfv{u}_{\mbox{\tiny N}}-\bfv{u}_{\mbox{\tiny L}}||$, is kept to be finite (non-vanishing) at the infinite Reynolds limit, where $\bfv{u}_{\mbox{\tiny N}}$ is the lower branch of NBW and $\bfv{u}_{\mbox{\tiny L}}$ is the laminar state.
Moreover, because the NBW is a derivative of HVS, which bifurcates from the laminar state via the secondary branch in the bifurcation diagram(\cite{sotita10}), NBW cannot directly connect to the laminar state even at the infinite $Re$ limit in a natural sense. %unless the bifurcation point of the derivative branch (NBW) becomes that of the primitive branch (HVS), which happens hardly in a natural sense, 
Therefore, NBW can ``not bifurcate from infinity''\cite{wan07}, contrary to the title of Ref.\cite{nag90}.
%, which was also previously pointed out by Wang {\it et al.} \cite{wan07}.

%\paragraph{\bf Behaviour of HVS at the high Reynolds}
In contrast with NBW, the lower branch of HVS may diminish with increasing $Re$.
%{\color{Blue} \st{This is because the bifurcation point of HVS (tertiary branch) from the secondary branch is quite close to the that of the secondary branch from the laminar state.} (Sotos-mou, I agree with you on the point that this sentence needs more explanation, but it says a vague conjecture. OK. Shall we delete this sentence.{\color{Red} Yes please delete it - it will take the paper over the 4 pages to explain - and it is a conjecture}}
%successively following the bifucation of the secondary branch from the laminar state.
%Thus, firstly
Firstly, 
 in the present study, by means of the continuation analysis with the aid of pseudo-spectral method, we traced the lower branch of HVS up to as high $Re$ as possible.
Several representative components of the lower branch of HVS are indicated in Fig.\ref{figB}. 
Referring to the first figure of Ref.\cite{wan07}, the amplitude of the streak component, $|u_0(y,z)-\bar{u}(y)|$, in the Fig.\ref{figB} is plotted explicitly for both states, NBW and HVS.
The lower branch of HVS algebraically looses all the harmonic amplitudes including its streak component with increasing $Re$, so that it asymptotically approaches the laminar state, which is distinct from the behaviour of NBW.
As it were, HVS could bifurcate from the laminar state at the infinite $Re$, while NBW could not.

%\section*{HVS and its manifold / BB}
\SEC{HVS on BB}
\begin{figure}
  \centerline{
    \includegraphics[angle=0,width=0.50\textwidth]{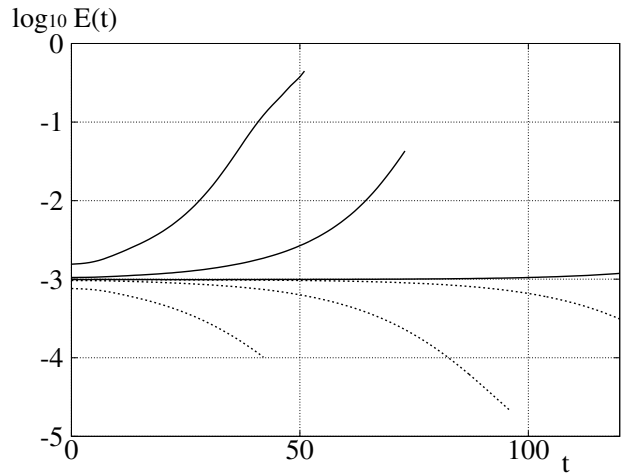}
  }
  \caption{
    $E(t)$ is half of norm of the wall-normal component of the perturbation at time $t$. %, which is nondimensionalised by $U$ and $h$.
    Trajectories starting around $\bfv{u}_{\mbox{\tiny H}}$ separates towards either the laminar (dotted) or the turbulent (solid) states due to a small  additive perturbation onto the initial state.
    Time development is carried out by imposing the reflection symmetry on the perturbation.
  }
  \label{figG}
\end{figure}
%\paragraph{\bf Lower branch of HVS exists on BB}
It remains an open question whether the lower branch of the more primitive solution than NBW, HVS, also lies on the BB at high $Re$.
In the present study, employing the {\it shooting method} \cite{ita01}, which was established originally as a tool in order to find out an unstable steady solution in a bistable system such as PCF, we specify the relation between the lower branch of HVS and the BB of PCF.
Thus, we followed the time-development of the flow starting from the lower branch of HVS, $\bfv{u}_{\mbox{\tiny H}}$, with a small perturbation, restricting the flow to satisfy the reflection symmetry in the spanwise direction.
 %\Skesu{calculated a number of time-development of flow state starting from the lower branch of HVS, $\bfv{u}_{\mbox{\tiny H}}$ with a small perturbation, where flow is restricted to satisfy the reflection symmetry to the spanwise direction.}
In Fig.\ref{figG}, the trajectories of the time-development of flow are depicted by a component of the velocity field, which is zero for the laminar regime and of order $1$ for the turbulent regime.
The closer the initial condition is to $\bfv{u}_{\mbox{\tiny H}}$ (the smaller the perturbation is), the longer transitional interval it takes for $\bfv{u}(t)$ to reach either turbulent or laminar regimes.
This means that HVS remains on the BB, and that it is an unstable solution even under the restricted dynamical system. 

%\paragraph{\bf BB approaches to the laminar}
Following these facts mentioned above, (1) the lower branch of HVS gets closer to the laminar with increasing $Re$ and (2) the lower branch of HVS stays on BB even at high $Re$, we can deduce the assumption that the minimum distance between the laminar state and BB might be representatively measured by $||\bfv{u}_{\mbox{\tiny H}}-\bfv{u}_{\mbox{\tiny L}}||$, which asymptotically vanishes with increase of $Re$.
Therefore, turbulent transition triggered by a small perturbation originating at HVS does not require an additional streamwise vortex component, even if the perturbation on the laminar state is infinitely small at the infinite $Re$.
Note that this is contrast to a conjecture on transition deduced from the asymptotic behaviour of NBW concluded in Ref.\cite{wan07}. 
In their conjecture, it is supposed that the formation of an $Re^{-1}$ updraft necessarily creates $O(1)$ streaks which in turn leads to turbulent transition via a certain nonlinear interaction between these modes.

%\section{one of manifolds of HVS towards NBW}
%\SEC{{\color{Red} The manifold of HVS (  (expanding towards) or (including ) NBW}}
\SEC{The manifold of HVS expanding towards NBW}
%\paragraph{\bf A plane on which LMN, HVS, NBW lie}
Additionally, we investigated one of manifolds of HVS, again employing the shooting method.
%\subsection*{HVS on BB}
%Following the earlier work by \cite{wan07}, which verified that the lower branch of NBW exists as an edge state on the BB of PCF, we investigate a structural aspect of the BB around HVS, the more primitive solution than NBW, by the shooting method.
%They showed that the lower branch of NBW has the only unstable manifold apart from the basin and that it is an attractor of the dynamical system restricted on the BB at high Reynolds number.
%It is also of interest to investigate a structural aspect of the BB around the more primitive exact solution, HVS.
Let us take a plane in state space on which the three distinct exact steady solutions of PCF, the laminar state, the lower branch of HVS, the lower branch of NBW, lie.
%The associated states are respectively depicted as $\bfv{u}_{\mbox{\tiny L}}$, $\bfv{u}_{\mbox{\tiny H}}$, and $\bfv{u}_{\mbox{\tiny N}}$, hereafter.
Using a couple of parameters $(a,b)$, we specify an arbitrary state $\bfv{u}_{(a,b)}$ on the plane, which is expressed as superposition of these solutions as follows
 \[\bfv{u}_{(a,b)}=b \bfv{u}_{\mbox{\tiny L}} + (1-b) \Bigl( (1-a) \bfv{u}_{\mbox{\tiny H}} + a \bfv{u}_{\mbox{\tiny N}}\Bigr) \ .\]
It should be noted that $\bfv{u}_{(a,b)}$ satisfies the incompressible condition, but is not necessarily equivalent to a steady state of the governing equation.
By adopting the state $\bfv{u}_{(a,b)}$ for a different set of $(a,b)$ as an initial state, $\bfv{u}(t=0)$, a number of trial calculations of the time development of the state $\bfv{u}(t)$ are carried out, which will provide part of the outline of the BB.

\begin{figure}
  \centerline{
   \includegraphics[angle=0,width=0.35\textwidth]{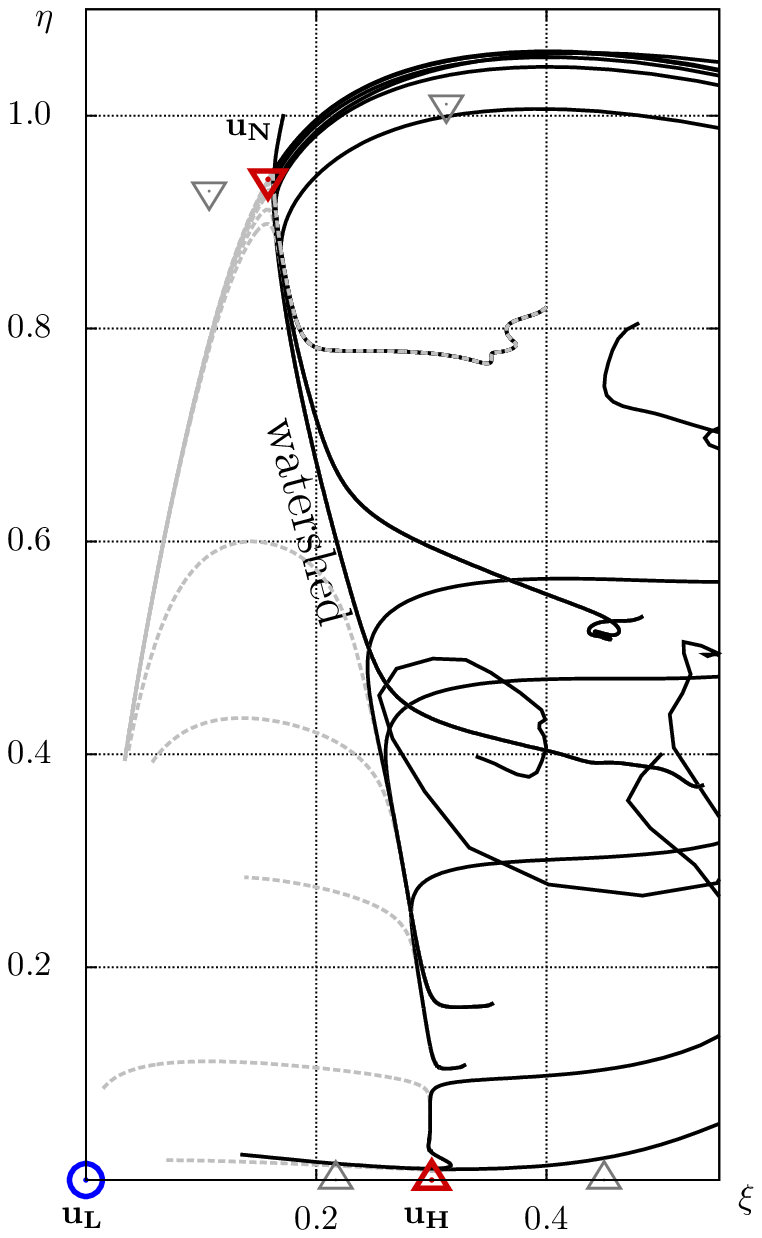}
  }
  \caption{ 
    %\color{Blue} 
    Trajectories starting from initial states, $\bfv{u}_{(a,b)}$ for a different set of parameters $(a,b)$, separate towards either the laminar $\bfv{u}_{\mbox{\tiny L}}$ (grey dashed curves) or the turbulent states (black solid curves), which are obtained at $Re=1000$.
    %An initial state consists of $\bfv{u}_{\mbox{\tiny H}}$($\triangle$) with a perturbation, which does not satisfy the reflection symmetry in the case of $y>0$.
    In the case of $\eta>0$, the initial state does not satisfy the reflection symmetry.
    %The final state, either laminar or turbulent, is determined only by a subtle difference of magnitude of additive perturbation to the initial state.
    Some of trajectories starting around $\bfv{u}_{\mbox{\tiny H}}$ at $Re=1000$ ({\color{red} $\triangle$}) with a particular set of $(a,b)$ trace a part of the BB as ``watershed'', which divides the whole space into the basins of the turbulent and laminar attractors.
    This BB is outlined by a hetero-clinic orbit\cite{hal09} connecting $\bfv{u}_{\mbox{\tiny H}}$ and $\bfv{u}_{\mbox{\tiny N}}$ at $Re=1000$ ({\color{red} $\bigtriangledown$}).
%, which is likely to be realized even at relatively high $Re$.
    For reference, $\bfv{u}_{\mbox{\tiny H}}$ and $\bfv{u}_{\mbox{\tiny N}}$ obtained at $Re=300, 10000$ are plotted as well (grey triangles), both of which approach towards $\xi=0$ with increasing $Re$.
    % All asymptotic behaviours indicate that the HVS instead of NBW can be referred to as the criterion of transition at high Reynolds numbers .
  }
  \label{figD}
\end{figure}

%\paragraph{\bf Projection of trajectories on x-y graph}
%\section*{PROJECTION}
{%\color{Blue} 
Here, let us plot $\bfv{u}(t)$ as a trajectory expressed by $(\xi(t),\eta(t),\zeta(t))$ in a three-dimensional space.
 % \Skesu{$(x(t),y(t),z(t))$ in a degenerated space with three dimensions.} 
The distance between any two different flow states $\bfv{u}_1$ and $\bfv{u}_2$ in the space is defined by the norm of the difference of the associated velocity field, $||\bfv{u}_1-\bfv{u}_2 ||$.
%,  \[ ||\bfv{u}_1,\bfv{u}_2||=\sqrt{\int_V (\bfv{u}_1-\bfv{u}_2)^2 dv} ,\] where $V$ is the total volume of one periodic box of channel.
%For instance, the distance of HVS from the laminar state is measured as $||\bfv{u}_{\mbox{\tiny H}}-\bfv{u}_{\mbox{\tiny L}}||$.
If one plots the laminar state at the origin, and the HVS on the $\xi$ axis $(\xi>0)$ with the distance, $||\bfv{u}_{\mbox{\tiny H}}-\bfv{u}_{\mbox{\tiny L}}||$,
 from the origin of the space,
 % in a natural sense, 
 then NBW may be plotted in the first quadrant on the $\xi-\eta$ plane so as to match the corresponding distances, $||\bfv{u}_{\mbox{\tiny N}}-\bfv{u}_{\mbox{\tiny L}}||$ and $||\bfv{u}_{\mbox{\tiny N}}-\bfv{u}_{\mbox{\tiny H}}||$, from the origin and HVS.
Thus, an arbitrary state $\bfv{u}$ can be plotted at a point in the space with the distances from the three exact states.
Fig.\ref{figD} is the projection of the trajectories, $\bfv{u}(t)$, onto the $\xi-\eta$ plane in the space.
Generally speaking, since the real state space consists of much higher dimensions, any state cannot maintain its identity in the projection.
%In addition, NBW bifurcated from HVS with breaking of a symmetry would allow another reflected image of NBW to be plotted at the half plane $y<0$.
}

%\paragraph{\bf Meaning of trajectories}
%\section*{UNSTABLE MANIFOLD of HVS HEADING towards NBW}
%With a variety of a couple of values $(a,b)$, a plenty of trial calculations are carried out to outline a part of the BB.
%In Fig.\ref{figD}, we plot only a could of trajectories starting from two different initial states using $b$ with a subtle difference ($a=0.04$ in both cases), on the plot.
%This implies that the BB must exist between two trajectories; one of the unstable manifolds of HVS heads towards NBW, which lies on the BB.
%the BB is emerged out of a tangle of trajectories in phase space, as if a ridge is outlined by a lot of small streams in mountainous area.
%This shooting method carried out at $Re=300$, but even if Reynolds number is increased $Re\sim 1000$, we could obtain similar behaviour as shown on the same plot.
Trajectories for a variety of a set values, $(a,b)$, obtained at $Re=1000$, are plotted in Fig.\ref{figD}.
Note that calculations with a variety of $b$ under fixed $a=0$ were demonstrated above (see Fig.\ref{figG}), which if plotted would be on the  $\xi$ axis in Fig.\ref{figD}.
In the case of $a=0$, the state $\bfv{u}(t)$ at a sufficient large $t$ decreases to $\bfv{u}_{\mbox{\tiny L}}$ for $b<0$, whereas it develops to some turbulent state for $b>0$.
% , which provides an evidence that the HVS is on the BB.
%`2. `Towards minimal perturbations in transitional plane Couette flow'', Yohann Duguet, Luca Brandt, and B. Robin J. Larsson, PHYSICAL REVIEW E 82, 026316 (2010)
In the figure, all trajectories starting from $\bfv{u}_{\mbox{\tiny H}}$ head eventually towards either $\bfv{u}_{\mbox{\tiny L}}$ or a turbulent state ($\xi^2+\eta^2>1$), but some of them pass by $\bfv{u}_{\mbox{\tiny N}}$.
A watershed, on which both $\bfv{u}_{\rm H}$ and $\bfv{u}_{\rm N}$ lie, emerges out of a tangle of trajectories, as if a ridge would been  outlined by a lot of small streams in \Skaku{a} mountainous area on a map.
The obtained watershed corresponds to the outline of the BB, which divides the whole space into the basins of the turbulent and laminar attractors, with the hetero-clinic orbit\cite{hal09} of these solutions constituting the BB. %, which is likely to be realized even at relatively high $Re$.
%The phase current on the BB comes from $\bfv{u}_{\rm H}$ to $\bfv{u}_{\rm N}$, which means that one of  unstable manifolds of $\bfv{u}_{\rm H}$ heads towards $\bfv{u}_{\rm N}$

%\section*{CONCLUDING REMARKS}
\SEC{Concluding remarks}
We will list briefly the facts obtained from the study on BB of PCF.
Firstly, the lower branch of HVS asymptotically approaches to the laminar state with increasing $Re$.
Secondly, the lower branch of HVS stays on the BB at high $Re$.
Thirdly, while one of unstable manifolds of HVS connects to the laminar state, another unstable manifold of HVS connects to the lower branch NBW, that is, a hetero-clinic orbit of these solutions constitutes the BB.
Additionally, NBW is a robust attractor in BB even at high $Re$, which is a conclusion given in Ref.\cite{wan07}.

From all four above facts we may deduce an ideal process of turbulent transition from the dynamical point of view.
In experiments of the turbulent transition, the magnitude of perturbation on the laminar is 
 %annihilated = deleted completely
 reduced % or shrunk
 as much as possible at the initial stage in the upstream.
In case that the flow experiences turbulent transition, though, the adopted smallest perturbation necessarily satisfies the reflection symmetry, which is inferred from the first and the second points.
From the third point, we may expect that the trajectory tends to approach towards NBW along the hetero-clinic orbit on BB.
In the downstream, NBW-type structures would be prevalent, while HVS is observed rarely, which is suggested from the fourth point.
This scenario gives us a clue to answer the question, why the vortex structure with the spanwise reflection symmetry is often observed at a transition stage in the turbulent boundary layer with a sufficiently large $Re$, 
 %where all the perturbation from the laminar state are kept to be as small as possible \Ikaku{at the upstream}, 
 where what is observed no longer depends on $Re$.
At the same time, this does not contradict that meandering streaky structure like NBW rather than HVS would be ubiquitous at downstream in fully developed turbulent shear flows (cf.\cite{sch10}).
Moreover the lower branch of HVS, rather than that of NBW, may give us a more practical criteria of laminar-turbulent transition of PCF at high $Re$. %, which provides the importance of HVS in the laminar-turbulent transition of PCF at high Reynolds numbers.
%One would tend to think that hairpin is dominat at less $Re$, but sinuous mode like NBW is dominant at hight $Re$.
%However, the scenario can explain that what is observed in the turbulent boundary layer with high $Re$ does not depend on 
Taking into account the optimal linear perturbation in PCF, we promote the comparison with the results of Ref. \cite{dug10}.%, which remains still open for the time being.

%\section*{Acknowledgement}
All of us would like to dedicate the paper to our families, whose continued support enabled us to research in the field of canonical shear flows.
In particular we are grateful to our fathers who over the years installed on us the enthusiasm to pursue vigorously the unknown in order to establish the truth.
%\st{T.I. is grateful for the financial support received from Aston University under the Visiting Scholars Fund Scheme and EPSRC (GR/S70593/01), which allowed him to spend time in the United Kingdom, while working on this project. S.C.G. is grateful to a Research Invitation from the Kansai University, which allowed him to spend time in Japan.} (Sotos-mou: Of course, I recognise that I owe very much on such past fund schemes {\color{Red} (please do not mention that - i enjoy working with you and i wanted you to come to UK this year, instead i will come - i am lucky to know you}, but I think even that relatively older acknowledgement should be also omitted now. What do you think ?{\color{Red} Yes please omit})
T.A. was supported by the European Union FP7 people Marie Curie International Incoming Fellowship grant: T2T-VDG F7-PEOPLE-2011-IIF 298891 .
This work has been also supported in part by KAKENHI (23760164) and by the Kansai University Special Research Fund 2012.

%\vspace{5mm}
%\ \ 

%\bibliographystyle{tsfp8}
%\bibliography{tsfp8}
\bibliographystyle{apsrev4-1}
%\begin{thebibliography}
\bibliography{prl05}
%\end{thebibliography}

\end{document}